\documentclass[aps,pra,twocolumn, noshowpacs, groupedaddress,nopacs,floatfix]{revtex4-1}
\usepackage{graphicx}
\usepackage{amssymb}
\usepackage{amsmath}

\newcommand{\bra}[1]{\ensuremath{\langle{#1}|}}
\newcommand{\ket}[1]{\ensuremath{|{#1}\rangle}}
\def\<{\langle}
\def\>{\rangle}

\begin{document}

\title{Nanophotonic quantum phase switch with a single atom}
\author{T. G. Tiecke$^{1,2}$}
\thanks{These authors contributed equally to this work}
\author{J. D. Thompson$^{1}$}
\thanks{These authors contributed equally to this work}
\author{N. P. de Leon$^{1,3}$}
\author{L. R. Liu$^1$}
\author{V. Vuleti\'{c}$^2$}\email{vuletic@mit.edu}
\author{M. D. Lukin$^1$}\email{lukin@fas.harvard.edu}
\affiliation{$^{1}$Department of Physics, Harvard University, Cambridge, MA 02138, USA}
\affiliation{$^{2}$Department of Physics and Research Laboratory of Electronics, Massachusetts Institute of Technology, Cambridge, MA 02139, USA}
\affiliation{$^{3}$Department of Chemistry and Chemical Biology, Harvard University, Cambridge, MA 02138, USA}

\maketitle

\textbf{In analogy to transistors in classical electronic circuits, a quantum optical switch is an important element of quantum circuits and quantum networks\cite{cirac97,kimble08,duan10}. Operated at the fundamental limit where a single quantum of light or matter controls another field or material system\cite{haroche06}, it may enable fascinating applications such as long-distance quantum communication\cite{briegel98}, distributed quantum information processing\cite{kimble08} and metrology\cite{komar13}, and the exploration of novel quantum states of matter\cite{carusotto13}.
Here, by strongly coupling a photon to a single atom trapped in the near field of a nanoscale photonic crystal cavity,  
we realize a system where a single atom switches the phase of a photon, and a single photon modifies the atom's phase. 
We experimentally demonstrate an atom-induced optical phase shift\cite{duan04} that is nonlinear at the two-photon level\cite{schuster08}, a photon number router that separates individual photons and photon pairs into different output modes\cite{aoki09}, and a single-photon switch where a single ``gate" photon controls the propagation of a subsequent probe field\cite{chen13,reiserer13}. These techniques pave the way towards  integrated quantum nanophotonic networks involving multiple atomic nodes connected by guided light.   
}
\begin{figure*}[htb]
\vspace{.0in} \centerline{\includegraphics[width=6in]{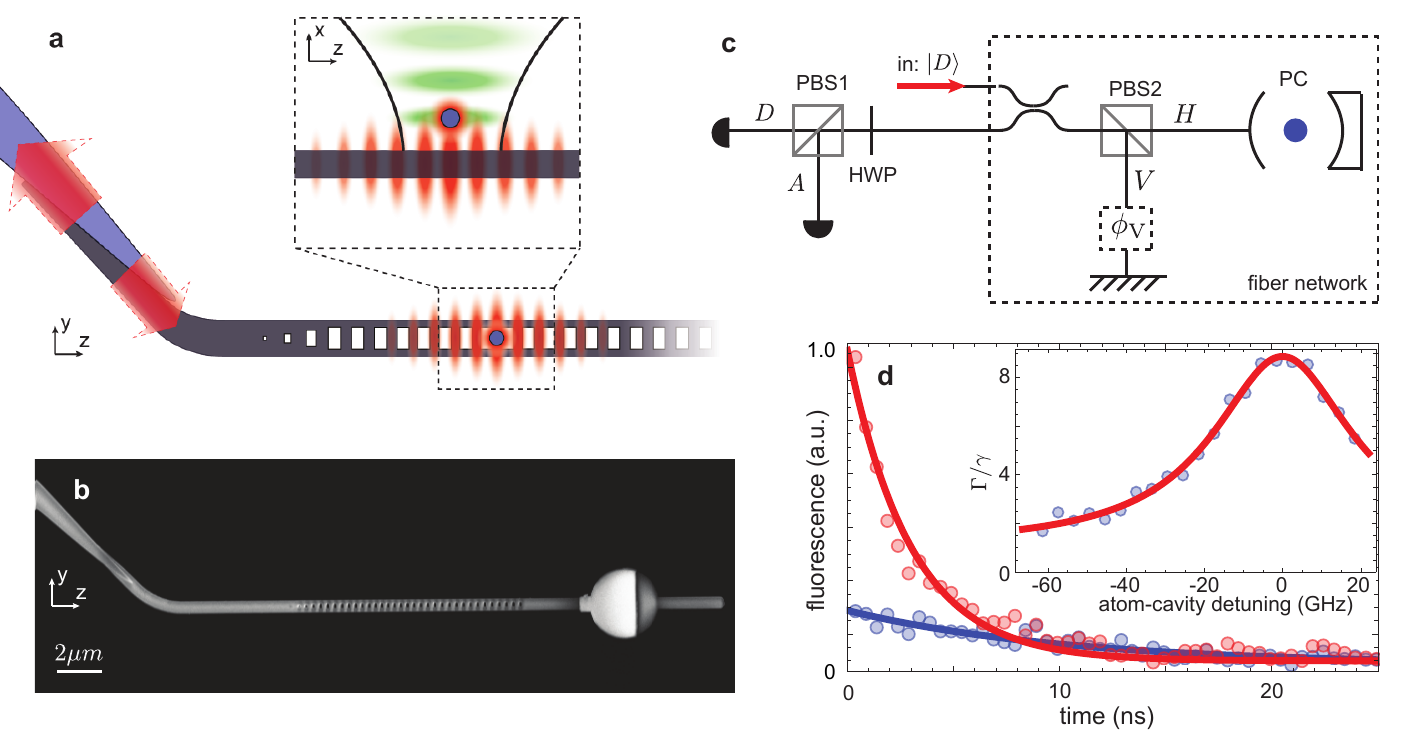}}
\vspace{.0in}\ 
\caption{\label{fig:fig1}
\textbf{Strong coupling of a trapped atom to a photonic crystal cavity.} \textbf{a.} A single $^{87}$Rb atom (blue circle) is trapped in the evanescent field (red) of a PC (gray). The PC is attached to a tapered optical fiber (blue), which provides mechanical support and an optical interface to the cavity. The tapered fiber-waveguide interface provides an adiabatic coupling of the fiber mode to the waveguide mode. The inset shows the one-dimensional trapping lattice (green), formed by the interference of an optical tweezer and its reflection from the PC. \textbf{b.} Scanning electron microscope (SEM) image of a single-sided PC. The pad on the right-hand side is used to thermally tune the cavity resonance by laser heating. \textbf{c.} The PC is integrated in a fiber-based polarization interferometer. A polarizing beamsplitter (PBS2) splits the $D$-polarized input field into an $H$-polarized arm containing the PC and a $V$-polarized arm with adjustable phase $\phi_\mathrm{V}$. Using a polarizing beamsplitter (PBS1) and half wave plate (HWP) the outgoing $D$ and $A$ polarizations are detected independently. \textbf{d.} Excited-state lifetime at an atom-cavity detuning of $0\,\mathrm{GHz}$ (red) and $-41\,\mathrm{GHz}$ (blue). The excited state lifetime is shortened to $\tau=\Gamma^{-1}=3.0(1)\,\mathrm{ns}$ from the free space value of $\gamma^{-1}=26\,\mathrm{ns}$, yielding a cooperativity $\eta=7.7 \pm 0.4$. The difference in the fluorescence signal at $t=0$ for the two detunings is consistent with the change in cavity detuning. The inset shows the enhancement of the atomic decay rate versus atom-cavity detuning.}
\end{figure*}

\begin{figure*}[htb]
\vspace{.0in} \centerline{\includegraphics[width=6in]{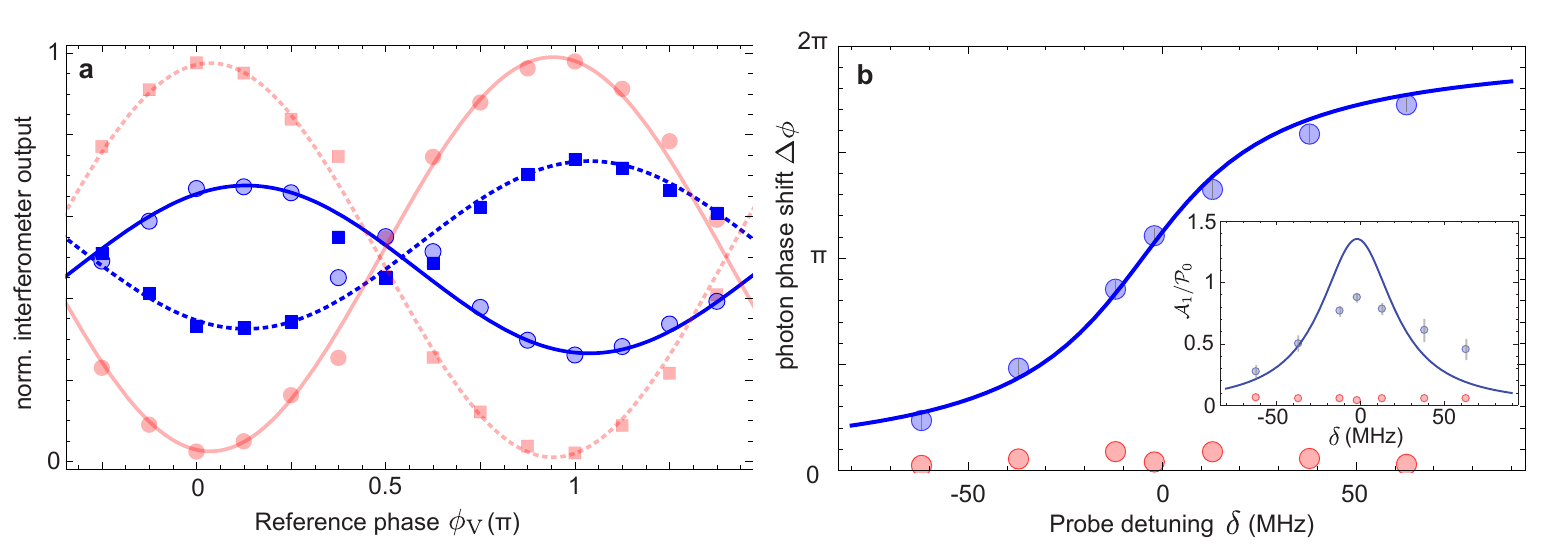}}
\vspace{.0in}\ \caption{\label{fig:fig2}\textbf{Photon phase shift  produced by a single atom.} \textbf{a.} Normalized interferometer output versus reference phase $\phi_\mathrm{V}$. The blue circles, blue squares, red circles, red squares correspond to $\mathcal{A}_1/\mathcal{P}_1$, $\mathcal{D}_1/\mathcal{P}_1$ (with atom) and $\mathcal{A}_0/\mathcal{P}_0$, $\mathcal{D}_0/\mathcal{P}_0$ (without atom) where $\mathcal{A}$ and $\mathcal{D}$ are the powers in the $A$ and $D$ output ports and $\mathcal{P}\equiv\mathcal{A}+\mathcal{D}$.   The measurement is performed near resonance ($\delta=-2\,\mathrm{MHz}$) and the lines are sinusoidal fits resulting in a phase shift of $(1.1\pm 0.1)\pi$. The maximum fringe visibility with and without an atom is $(44 \pm 2)\%$ and $(97 \pm 1)\%$, respectively. \textbf{b.} Measured phase shift versus detuning in the presence (blue) or absence (red) of an atom. The curve includes cavity losses in Eq. \ref{eq:reflect} (see SI), and corresponds to a cooperativity of $\eta=7.7$ and a small ($5\,\mathrm{MHz}$) offset from the free-space resonance. 
The inset shows $\mathcal{A}_1/\mathcal{P}_0$ at $\phi_\mathrm{V}=\pi$. The solid line is the expected value for the same model parameters as in the main figure. The expected increase in reflectivity in the presence of an atom ($\mathcal{P}_1/\mathcal{P}_0 > 1$) arises because the atom reduces the field amplitude in the lossy cavity (see SI). In our experiment we observe $\mathcal{P}_1/\mathcal{P}_0 \simeq 1.2$. The error bars reflect $\pm 1 \sigma$ statistical uncertainty.}
\end{figure*}

A quantum optical switch\cite{oshea13,volz12,kim13,chen13,chang07} is challenging to implement because the interaction between individual photons and atoms is generally very weak. Cavity quantum electrodynamics (cavity QED), where a photon is confined to a small spatial region and made to interact strongly with an atom, is a promising approach to overcome this challenge\cite{haroche06}. Over the last two decades, cavity QED has enabled advances in the control of microwave\cite{schuster07,gleyzes07,deleglise08} and optical fields\cite{turchette95,fushman08,aoki06,ritter12,oshea13}. While integrated circuits with strong coupling of microwave photons to superconducting qubits are currently being developed\cite{devoret13}, 
a scalable path to integrated quantum circuits involving coherent qubits coupled via optical photons has yet to emerge.

Our experimental approach, illustrated in Figure 1a, makes use of a single atom trapped in the near field of a nanoscale photonic crystal (PC) cavity that is attached to an optical fiber taper\cite{thompson13b}. The tight confinement of the optical mode to a volume $V \sim 0.4\,\lambda^3$, below the scale of the optical wavelength $\lambda$, results in strong atom-photon interactions for an atom sufficiently close to the surface of the cavity. The atom is trapped at about 200 nm from the surface in an optical lattice formed by the interference of an optical tweezer and its reflection from the side of the cavity (see Methods Summary, SI and Fig. 1a,b). Compared to transient coupling of unconfined atoms\cite{aoki06,oshea13}, trapping an atom allows for experiments exploiting long atomic coherence times, and enables scaling to quantum circuits with multiple atoms.

We use a one-sided optical cavity with a single port for both input and output\cite{duan04}. In the absence of intracavity loss, photons incident on the cavity are always reflected. However, a single, strongly-coupled atom changes the phase of the reflected photons by $\pi$ compared to an empty cavity. More specifically, in the limit of low incident intensity, the amplitude reflection coefficient of the atom-cavity system is given by\cite{wachs06}:
\begin{align}
r_\mathrm{c}(\eta) & = \frac{(\eta-1)\gamma +2i\delta}{(\eta+1)\gamma -2i\delta}
\label{eq:reflect}
\end{align}
where $\eta=(2g)^2/(\kappa \gamma)$ is the cooperativity, $2g$ is the single photon Rabi frequency, $\delta$ is the atom-photon detuning, and the cavity is taken to be resonant with the driving laser. In our apparatus, the cavity intensity and atomic population decay rates are given by $\kappa=2\pi \times 25\,\mathrm{GHz}$ and $\gamma=2\pi \times 6\,\mathrm{MHz}$, respectively. The reflection coefficient in Eq. \ref{eq:reflect} changes sign depending on the presence ($\eta > 1$) or absence ($\eta=0$) of a strongly-coupled atom. 
If the atom is prepared in a superposition of internal states, one of which does not couple to the cavity mode
(\emph{e.g.} another hyperfine atomic sublevel),  the phase of the atomic superposition is switched by $\pi$ upon the reflection of a single photon. 
By also adding an auxiliary photon mode  that does not enter the cavity (\emph{e.g.}, an orthogonal polarization), this operation can be used to realize the Duan-Kimble scheme for a controlled-phase gate between an atomic and a photonic quantum bit\cite{duan04}. The property of the atom-cavity system that a single photon and a single atom can switch each other's phase by $\pi$ is the key feature of this work. 

We quantify the single-atom cooperativity $\eta$ by measuring the lifetime $\tau$ of the atomic excited state when it is coupled to the cavity. We excite the atom with a short $(3\,\mathrm{ns})$ pulse of light co-propagating with the optical trap and resonant with the $|5S_{1/2},F=2\rangle \rightarrow |5P_{3/2},F'=3\rangle$ transition (near $780\,\mathrm{nm}$). The atomic fluorescence is collected through the cavity to determine the reduced excited-state lifetime $\tau=\Gamma^{-1}$, as shown in Fig. 1d, which yields the cooperativity $\eta=(\Gamma - \gamma)/\gamma$. Fitting a single exponential decay gives $\tau = (3.0 \pm 0.1)\,\mathrm{ns}$, corresponding to $\eta=7.7 \pm 0.3$ and a single-photon Rabi frequency of $2g = 2\pi \times (1.09 \pm 0.03)\,\mathrm{GHz}$. 

\begin{figure*}[p]
\vspace{.0in} \centerline{\includegraphics[width=6in]{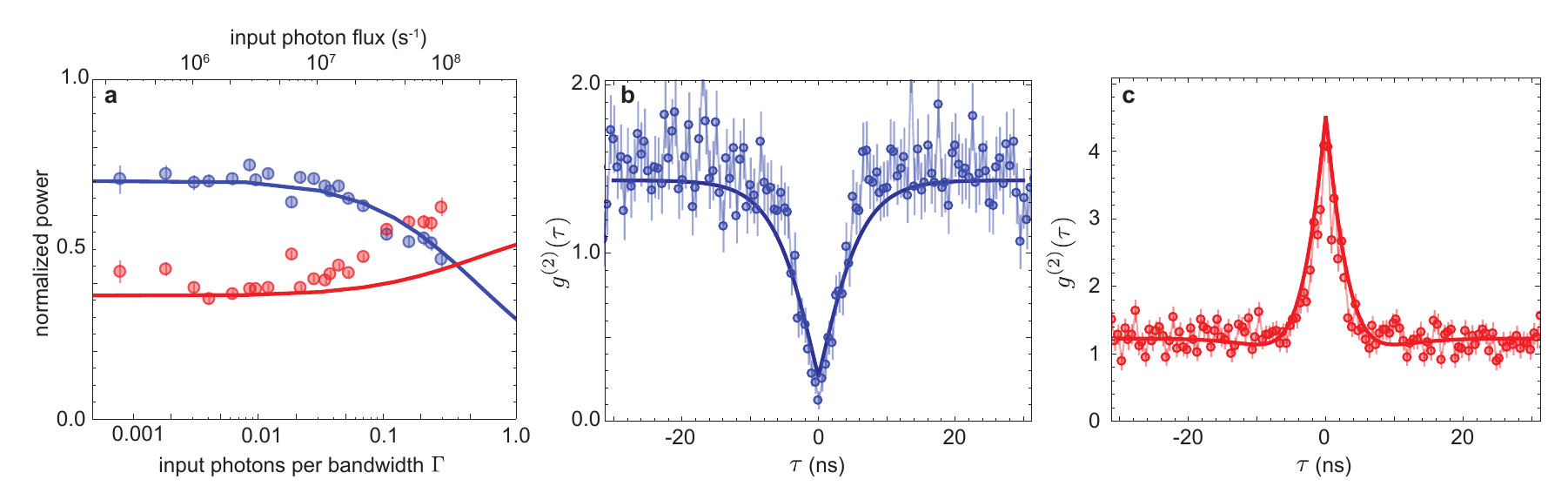}}
\vspace{.0in}\ \caption{\label{fig:fig3} \textbf{Quantum nonlinear optics with the atom-PC system} \textbf{a.} Interferometer output as a function of the photon rate incident on the interferometer. The outputs $\mathcal{A}_1/\mathcal{P}_0$ (blue) and $\mathcal{D}_1/\mathcal{P}_0$ (red) are normalized to the case without an atom. The incident photon rate is normalized to the enhanced atomic decay rate $\Gamma = (\eta+1) \gamma$. The interferometer is tuned such that port $A$ is dark in the absence of the atom and the output in port $A$ starts to saturate at a rate below one photon per bandwidth $\Gamma$. Unlike the data in Figure 2 and 4 these measurements were performed in the presence of the dipole trap which reduces  $\mathcal{A}_1/\mathcal{P}_1$ at low driving intensities (see SI). \textbf{b-c.}  Photon-photon correlation functions $g^{(2)}(\tau)$ for the $A$ (\textbf{b}) and $D$ (\textbf{c}) ports. Port $A$ shows clear anti-bunching with $g_R^{(2)}(0)=0.12(5)$, while port $D$  exhibits a strong bunching of $g_T^{(2)}(0)=4.1(2)$. The solid lines in figure $\mathbf{a}$-$\mathbf{c}$ are obtained from a model including inhomogeneous light-shift broadening arising from the dipole trap (see SI). The error bars reflect $\pm 1 \sigma$ statistical uncertainty.} 
\end{figure*}

\begin{figure*}[p]
\vspace{.0in} \centerline{\includegraphics[width=6in]{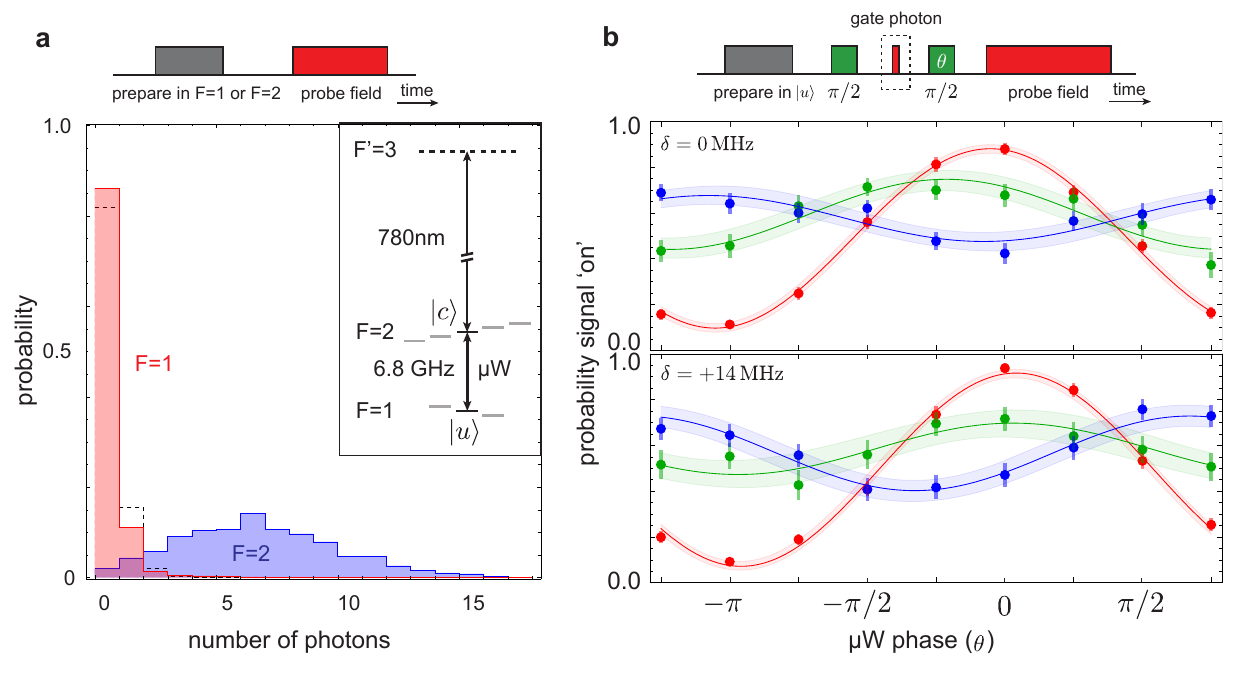}}
\vspace{.0in}\ \caption{\label{fig:fig4}\textbf{Realization of the quantum phase switch} \textbf{a.} Number of probe photons detected in port $A$ as a function of the internal atomic state. If the atom is in the $F=2$ manifold the switch field is ``on'', thereby routing $\bar n_A = 6.2$ photons to port $A$. If the atom is absent (dashed line) or in the $F=1$ manifold, $\bar n_A = 0.2$. The input photon number is the same in all cases, with a peak rate much smaller than $\Gamma$. The separation between the two distributions allows the switch states to be distinguished with 95\% average fidelity. The inset shows the relevant levels for the quantum switch. The laser is tuned to the $F=2$ to $F'=3$ transition, and couples only to $|c\rangle$. \textbf{b.} (top) The switch sequence (see text).
(bottom) The probability $P_{on}$ of finding the switch ``on'', as a function of the phase $\theta$ of the second microwave pulse ($\delta=0$ (top panel) and $\delta=2\pi \times 14\,\mathrm{MHz}$ (bottom panel)). $P_{on}$ is shown in several cases: without a gate field ($P_{on}^{0}$, red); and with a gate field, both with ($P_{on}^{1}$, blue) and without ($P_{on}^{uc}$, green) conditioning on the detection of a reflected photon. The error bars reflect $\pm 1 \sigma$ statistical uncertainty in the data, while the shaded region shows the range of curves with fit parameters within $1 \sigma$ of the best fit.} 
\end{figure*}

To probe the optical phase shift resulting from the atom-photon interaction, we integrate the cavity into a fiber-based polarization interferometer, which converts phase shifts into polarization rotations (Figure 1c). The $H$-polarized arm of the interferometer contains the cavity, while the $V$-polarized arm is used as a phase reference.  For an input photon state $\ket{\psi_{in}}$ in the polarization basis $\{\ket{H},\ket{V}\}$, the state exiting the interferometer is given by $\mathcal{R} |\psi_{in}\rangle$, where $\mathcal{R}  \equiv r_{\mathrm{V}} e^{i \phi_{\mathrm{V}}}|V\rangle\langle V| + r_c(\eta) |H\rangle \langle H|$ and $r_{\mathrm{V}}$, $\phi_{\mathrm{V}}$ are the amplitude and phase of the reflection of the reference arm. We choose the reflectivity $r_V$ of the reference arm to match that of the empty (lossy) cavity (see SI), such that in the absence of an atom, the light emerges in the incident polarization state $|D\rangle\equiv(|V\rangle+|H\rangle)/\sqrt{2}$. In the presence of an atom, for $\phi_V=0$ and $\eta \gg 1$, input light exits the interferometer predominantly with the orthogonal polarization $|A\rangle\equiv(|V\rangle-|H\rangle)/\sqrt{2}$. 

Figure 2a demonstrates the optical phase shift arising from an atom coupled to the cavity. A weak $D$-polarized probe field is applied at the interferometer input, and the output power in the $A$ and $D$ ports is recorded as a function of the reference phase $\phi_V$. The phase of the reflected light is shifted by $(1.1 \pm 0.1)\pi$ relative to the case with no atom, and the visibility of the oscillation with $\phi_V$ is $(44 \pm 2) \%$ and $(39 \pm 2) \%$ in the $A$ and $D$ ports, respectively. By repeating this measurement for a range of atom-photon detunings $\delta$, we observe a $2\pi$ change in the reflection phase across the atomic resonance (Figure 2b), in agreement with Eq. (\ref{eq:reflect}). For the data presented, the events where an atom was not present in the cavity (\emph{e.g.} by escape from the trap) were excluded. The remaining contributions to the reduced fringe visibility are imperfect balancing of the interferometer ($\sim5\%$), atomic saturation effects ($\sim10\%$), state-changing scattering processes that leave the atom in a different final state and therefore reveal which-path information in the interferometer ($\sim20\%$) and thermal motion of the atom ($\sim20\%$) (see SI). 

The saturation behavior of the atom-cavity system is examined in Figure 3a, which shows the fraction of the output power in the $A$ and $D$ ports as a function of the input power. We set the reference phase $\phi_\mathrm{V}\simeq0$ such that the $A$ port is dark in the absence of the atom.
The distribution of the output is power-indepedent for low input powers, as expected for a linear system. At higher powers, the atomic response saturates and the output fraction at the $A$ port decreases. The saturation becomes evident when the input photon rate approaches the enhanced excited state decay rate $\Gamma$, in agreement with theoretical predictions (see SI). This nonlinearity results in different reflection phases for single photons and photon pairs. In a Hanbury-Brown-Twiss experiment, we measure the photon-photon correlation functions $g^{(2)}(\tau)$ at low input power. We observe strong anti-bunching of $g_A^{(2)}(0)=0.12(5)$ and bunching of $g_D^{(2)}(0)=4.1(2)$ in the $A$ and $D$ ports respectively, indicating that the atom-cavity system acts as an effective photon router by sending single photons into output $A$ and photon pairs into output $D$\cite{witthaut12}.

To realize a quantum switch where the state of a single atom controls the propagation of many probe photons, we use two atomic hyperfine states, $|c\rangle \equiv |F=2,m_F=0\rangle$ and $|u\rangle \equiv |F=1,m_F=0\rangle$ (see Figure 4a)  which can be coherently manipulated with microwaves. 
While the atom-photon interaction strength is similar for all of the sublevels in a given hyperfine manifold, the $F=1$ levels (including $|u\rangle$) are effectively uncoupled because the probe is far-detuned from all optical transitions originating from this level. In Fig. 4b, we show the output signal at the $A$ port for a $D$-polarized probe field with an atom prepared in $F=1$ or $F=2$. The switch is ``on" and the input light goes mostly to the $A$ port when the atom is in $F=2$, while the switch is ``off" and the $A$ port is dark when the atom is in $F=1$. We estimate that up to $\bar n_A \simeq 75$ photons could be transmitted to the $A$ port in the ``on" state before the atom is optically pumped out of the $F=2$ manifold. In the experiments shown in Figure 4, a smaller number of photons ($\bar n_A = 6.2$) was used to increase the rate of data acquisition by allowing a greater number of measurements with the same atom. This photon number allows us to distinguish the switch state with an average fidelity of 95\%.

As the effect of an atom on a photon and that of a photon on an atom are complementary, it follows from Eq. \ref{eq:reflect} that a single photon can shift the phase of the coupled state $\ket{c}$ by $\pi$. This phase shift can be converted into a flipping of the atomic switch, $\ket{c} \leftrightarrow \ket{u}$, using an atomic Ramsey interferometer\cite{gleyzes07}. An atom is first prepared in the $|u\rangle$ state via optical pumping and rotated to the superposition $(|u\rangle + |c\rangle)/\sqrt{2}$ by a microwave $\pi/2$ pulse (see SI). A single $H$-polarized ``gate" photon flips the atomic superposition to $(|u\rangle - |c\rangle)/\sqrt{2}$. As reflection of the gate photon does not reveal the atomic state, the atomic superposition is not destroyed. Finally, a second microwave $\pi/2$ pulse rotates the atomic state to $|c\rangle$ or $|u\rangle$ depending on the presence or absence of the gate photon, leaving the switch on (atom in $\ket{c}$) or off (atom in $\ket{u}$). A similar technique was recently explored for nondestructive photon detection in a Fabry-Perot cavity\cite{reiserer13}.

In our measurement, we mimic the action of a single gate photon by applying a weak coherent field with $\bar{n} \approx 0.6$ incident photons and measuring the probe transmission conditioned on the detection of a reflected gate photon at either interferometer output. Fig. 4c shows the probability $P_{on}$ to find the switch in an ``on'' state as a function of the phase of the second microwave pulse. The dependence of $P_{on}$ on the microwave phase when a reflected gate photon is detected shows that the superposition phase is shifted by $(0.98 \pm 0.07) \pi$. The atomic coherence is reduced but not destroyed. The absence of a phase shift in the unconditioned data (green curve in Fig. 4c) confirms that the switch is toggled by a single photon. The phase shift depends on the gate photon detuning: tuning the laser to $\delta=2\pi \times 14\,\mathrm{MHz}$ results in a phase shift of $(0.63 \pm 0.15) \pi$, in good agreement with the detuning dependence of the photon phase shift (Figure 2b). 

For an optimally chosen phase of the second microwave pulse, we find that the switch is in the ``on'' state with probability $P_{on}^{1} = 0.64 \pm 0.04$ if a gate photon is detected, $P_{on}^{0} = 0.11 \pm 0.01$ if no gate field is applied, and $P_{on}^{uc} = 0.46 \pm 0.06$ without conditioning on single photon detection. The finite $P_{on}^0>0$ without a gate field arises from imperfect atomic state preparation and readout fidelity (see SI). $P_{on}^1$ is also affected by the finite probability for the gate field to contain two photons, of which only one is detected. This results in a decrease (increase) of $P_{on}^{1}$ ($P_{on}^{uc}$) by about $20\%$ in a way that is consistent with our measurements (see Methods Summary and SI). 
We attribute the $8\,\%$ positive offset in $P_{on}^1$ and $P_{on}^{uc}$ to spontaneous scattering events of the gate photon, which cause atomic transitions to a final state other than $|c\rangle$ within the $F=2$ manifold. 
Lastly, we estimate that fluctuations in $\eta$ arising from thermal motion do not change $P_{on}^1$ by more than 10\%, since the atom-photon interaction scheme used here\cite{duan04} is inherently robust to variations in $\eta$ for $\eta \gg 1$. 
The imperfect fringe visibility in Figure 2 and 4, due to the technical imperfections discussed above, can be improved by better atomic state preparation, alignment of the cavity polarization with the magnetic field defining the quantization axis, and improved atom localization. The fringe visibility does not directly depend on the cooperativity and absent technical imperfections, perfect fringe visibility should be achievable; however, the probability of gate photon loss is reduced as the cooperativity increases (see SI).

Our experiments open the door to a number of intriguing applications. For instance, efficient atom-photon entanglement for quantum networks can be generated by reflecting a single photon from an atom prepared in a superposition state. The quantum phase switch also allows for quantum non-demolition measurements of optical photons\cite{volz11,reiserer13}. With an improved collection efficiency of light from the PC cavity and reduced cavity losses, it should be possible to make high-fidelity non-demolition measurements of optical photon number parity to create non-classical ``cat"-like states\cite{wang05}, with possible applications to state purification and error correction. Most importantly, the scalable nature of both nanofabrication and atomic trapping allow for extensions of this work to complex integrated networks with multiple atoms and photons. 

\subsection{Methods Summary}
We begin our experiments by loading a single $^{87}$Rb atom from a magneto-optical trap into a tightly focused optical dipole trap. After a period of Raman sideband cooling\cite{thompson13a} to localize the atom in the trapping potential, we translate the optical dipole trap to the PC cavity, where the interference of the dipole trap light with its reflection from the PC forms an intensity maximum that confines the atom at a distance of about 200 nm from the surface of the PC\cite{thompson13b} (Fig. 1a,b). The success probability of loading an atom near the PC is $>\,90\%$. We modulate the dipole trap with full contrast at $5\,\mathrm{MHz}$ to interrogate the trapped atom at instances that the light-shift is negligible. 

\subsection{Acknowledgements} We acknowledge T. Peyronel, A. Kubanek, A. Zibrov for helpful discussions and experimental assistance. Financial support was provided by the NSF, the Center for Ultracold Atoms, the Natural Sciences and Engineering Research Council of Canada, the Air Force Office of Scientific Research Multidisciplinary University Research Initiative and the Packard Foundation. JDT acknowledges support from the Fannie and John Hertz Foundation and the NSF Graduate Research Fellowship Program. This work was performed in part at the Center for Nanoscale Systems (CNS), a member of the National Nanotechnology Infrastructure Network (NNIN), which is supported by the National Science Foundation under NSF award no. ECS-0335765. CNS is part of Harvard University.

\pagebreak
\widetext

\renewcommand{\thefigure}{S\arabic{figure}}
\renewcommand{\theequation}{S\arabic{equation}}
\makeatletter
\renewcommand*{\@biblabel}[1]{[S\hfill#1]}
\renewcommand*{\@cite}[1]{[S#1]}
\makeatother

\section*{Supplementary material}
\section{Experimental setup}
\subsection{Apparatus}
Our setup is described in detail in Ref. \cite{thompson13a,thompson13b}, and is only briefly reviewed here. The apparatus consists of an ultra-high vacuum (UHV) system with a $^{87}$Rb MOT. We trap single atoms in a tightly focused scanning optical tweezer (waist $w_0=0.9\,\mathrm{\mu m}$, wavelength $\lambda = 815\, \mathrm{nm}$, trap depth $U_0=1.0\,\mathrm{mK}$), which is formed at the focus of an aspheric lens (Thorlabs 352240). After loading an atom from the MOT and performing Raman sideband cooling  \cite{thompson13a} to better localize the atom, we increase the tweezer depth to $U_0=2.1\,\mathrm{mK}$ and translate the tweezer to the photonic crystal cavity, about $40\,\mathrm{\mu m}$ away. At its final position the optical dipole trap is formed by the interference of the optical tweezer with its reflection from the PC, creating a 1D optical lattice. Based on numerical simulations we estimate the closest lattice site to be $180\,\mathrm{nm}$ away from the surface, with a maximum light shift of $\sim 4\,\mathrm{mK}$. The trap depth is smaller than this maximum value because of a finite light intensity at the surface of the PC and additional surface forces. From measurements of the atom-cavity coupling, we infer that the loading procedure succeeds more than 90\% of the time. The lifetime of the atom in the tweezer is about $0.25\,\mathrm{s}$ near the photonic crystal, which is shorter than the lifetime in the tweezer in free space ($6\,\mathrm{s}$). To ensure relative position stability of the tweezer and the PC we periodically measure the position of the PC and adjust our coordinate system for observed drifts. The PC position is determined by inserting $815\,\mathrm{nm}$ light into the PC and detecting the emitted light through the optical tweezer path. By taking 5 images in different focal planes the PC position is determined in 3D.

The finite temperature of the atom leads to time-varying light shifts of the optical transition in the presence of the dipole trap \cite{thompson13b}. In order to suppress this effect in the measurements presented in Figures 2 and 4, we modulate the dipole trap intensity with full contrast at $5\,\mathrm{MHz}$ and probe the atom-photon interaction only when the intensity is nearly zero. Since this modulation is much faster than the highest trap frequency ($710\,\mathrm{kHz}$), the atom experiences a time-averaged potential and the trapping potential is well-described by the potential averaged over one modulation period, as explored in time-orbiting potentials for ultra cold atoms and RF Paul traps for ions. For modulation frequencies above $4\,\mathrm{MHz}$ we observe no reduction of the trap lifetime compared to the unmodulated trap. The modulation is produced by dividing the dipole trap beam into two paths, shifting with two coherently driven acousto-optic modulators (AOM) detuned by the desired modulation frequency, and recombining the two AOM outputs into a single-mode fiber. When applying this modulation we observe the optical transition frequency to be within $5\,\mathrm{MHz}$ of its free space value. In an unmodulated trap of the same average intensity it is shifted by $\sim 120\,\mathrm{MHz}$.

For the measurements in Figures 2 and 4, both the probe and gate pulses consist of a train of Gaussian pulses with a FWHM of $24\,\mathrm{ns}$. These pulses are generated by a fiber-based electro-optic modulator (Jenoptik AM 830) driven by an arbitrary waveform generator (Agilent 33250A). Synchronization with the dipole trap modulation is achieved by triggering the pulse train with a low-jitter delay generator (SRS DG645) from a photodiode which monitors the dipole trap power directly.

All measurements were performed with single photon counting modules (Perkin Elmer SPCM-AQRH), recorded using a PicoHarp 300 time-correlated single-photon counting system.

\subsection{Polarization interferometer}
In section \ref{sect:theory} we give a detailed theoretical description of the input and output modes of the interferometer. Here, we discuss the experimental implementation.
\subsubsection{Experimental implementation}
\begin{figure}[tpb]
\begin{center}
\includegraphics[width=1.0 \textwidth]{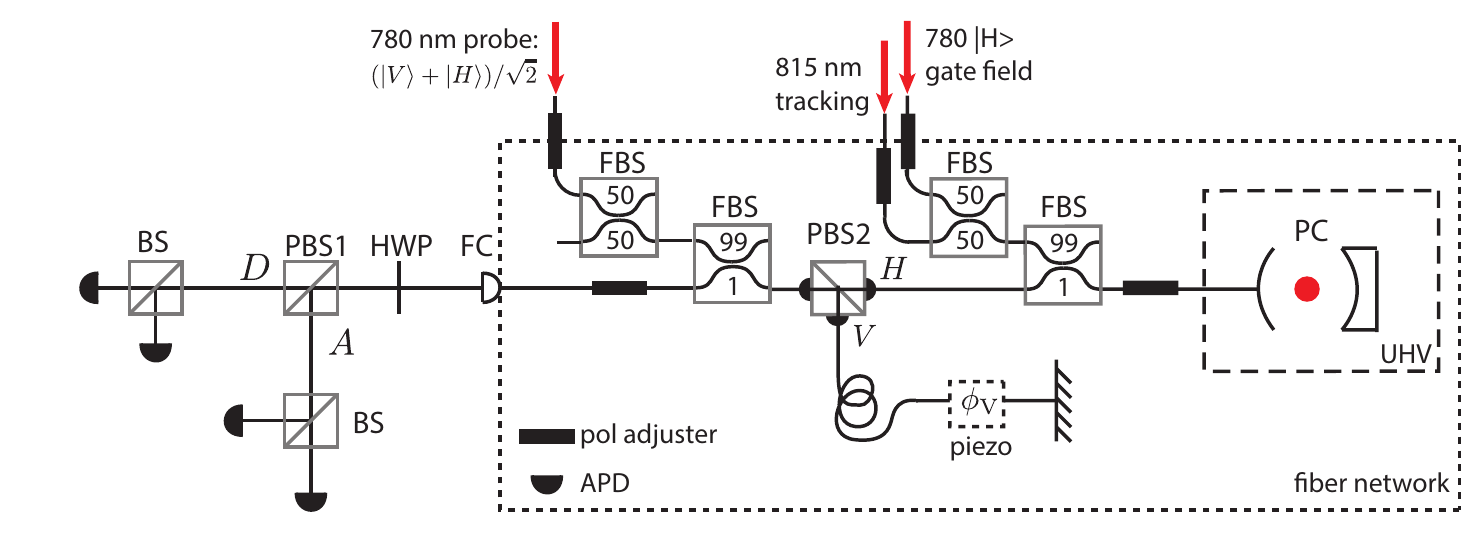}
\caption{\label{fig:setup}Schematic drawing of the fiber based polarization interferometer. All components within the dashed line are fiber-based. PBS1 and PBS2 are free-space and a fiber-based polarization beamsplitters, respectively. BS denotes 50/50 beamsplitters and HWP and FC are a $\lambda/2$-plate and a fiber coupler. In both the $D$ and $A$ ports of the interferometer a pair of detectors is used for photon-photon correlation measurements. The fiber beamsplitters (FBS) are labeled with their coupling ratios. The two $780\,\mathrm{nm}$ input fields are used for coupling to the atom and for stabilizing the cavity and interferometer, while the $815\,\mathrm{nm}$ field is used for stabilization of the device position. }
\end{center}
\end{figure}

The PC is mounted inside the UHV system attached to a tapered optical fiber. The fiber is guided out of the UHV system through a fiber feedthrough and integrated into a fiber based interferometer (see Figure \ref{fig:setup}). All fiber-fiber connections are fusion spliced to ensure high coupling efficiencies and we achieve a total efficiency from the free space fiber coupler to the tapered fiber of $78\%$, mostly limited by PBS2. The fiber of the $|V\rangle$ polarized reference arm is glued to the side of a piezo stack, which allows for tuning $\phi_\mathrm{V}$ over many tens of $\pi$. We adjust the polarization of the various arms by means of fiber polarization controllers. We find that optimizing the polarization controllers once every few weeks is sufficient for stable operation of the interferometer.

The path length of the two arms of the interferometer are adjusted to be within several mm of each other, so that the free spectral range of the interferometer is large ($> 30\,\textrm{GHz}$) compared to the range of frequencies used to probe the atom.

Thermal effects cause fluctuations of the relative phase of the two interferometer arms. We compensate for these drifts by stabilizing $\phi_V$ such that the power in the $A$ port is minimized in the absence of an atom. In order to obtain an error signal for the stabilization we send a $780\,\mathrm{nm}$ probe beam through the interferometer while dithering $\phi_V$. We use a field programmable gate array (FPGA) to implement lock-in detection of the modulated probe reflection and apply feedback to $\phi_V$. This feedback is applied during the Raman cooling sequence (which lasts $\sim 150\,\mathrm{ms}$). 

\subsection{Photonic crystal design and fabrication}
The PC cavities are fabricated using electron beam lithography and reactive ion etching, as described previously in \cite{thompson13b}. The cavities used in this work are formed from waveguides with a cross section of $500\,\mathrm{nm}$ by $175\,\mathrm{nm}$, and are patterned with rectangular holes of size $225\,\mathrm{nm}$ by $126\,\mathrm{nm}$. The pitch of the holes is $280\,\mathrm{nm}$ in the center of the cavity, and gradually increases to $315\,\mathrm{nm}$ on either end of the cavity. To make the cavity single-sided, there are 5 extra holes on the side of the cavity opposite the fiber, which increases the reflectivity of this mirror by a factor greater than 10.

To enable the cavities to be heated with a laser for thermal tuning of the transition frequency, a pad is formed on the waveguide (as visible in Figure 1c) and coated with amorphous silicon. Depositing this material is the final step in the fabrication process, to allow the absorbing material to be chosen independently of its compatibility with the strong acids and bases used for undercutting and cleaning the waveguides. This is accomplished by using a patterned silicon nitride membrane as a stencil for electron beam evaporation of the absorbing material. The membrane is held several microns over the top of the cavities with a spacer, and aligned to deposit the absorbing material on the pads without contaminating the cavities.

After fabrication, an array of cavities is characterized with a tapered fiber probe. Using the linewidth and the amount of power reflected at the cavity resonance frequency, we can extract both the decay rate into the waveguide and the decay rate into other modes that we do not collect. In the set of cavities fabricated for this experiment, the decay rates into other modes ranged from $\kappa_{sc}=4-15\,\mathrm{GHz}$, corresponding to loss-limited quality factors of about 30,000 - 100,000. The waveguide decay rate was fixed by the fabrication parameters to be $\kappa_{wg}=20\,\mathrm{GHz}$, ensuring that all cavities are over-coupled.

Finally, a single PC is selected, removed from the substrate, attached to a tapered optical fiber  and inserted into the UHV chamber. The fiber-waveguide coupling efficiency is $62\,\%$.

\subsection{Interferometer and PC characterization}

\begin{figure}[tpb]
\begin{center}
\includegraphics[width=0.5 \textwidth]{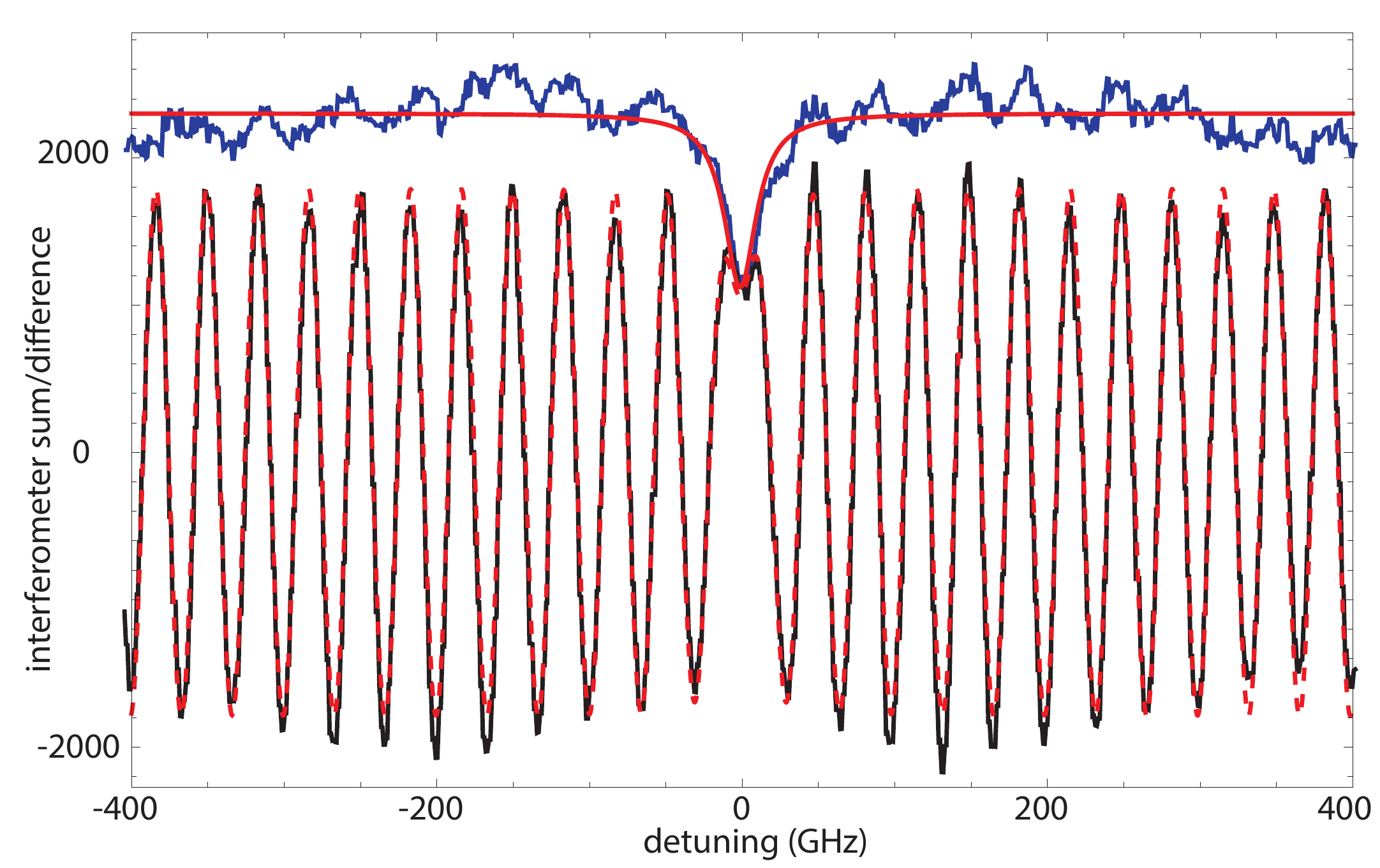}
\caption{\label{fig:interfFSR} Characterization of the interferometer and PC. The solid blue and black curves are the measured  sum and difference of the two interferometer ports respectively. The red solid and dashed lines are the fit to Eq. \ref{eq:DpA} and Eq. \ref{eq:DmA} respectively.}
\end{center}
\end{figure}

We characterize the cavity and interferometer using a New Focus Velocity TLB-6712 tunable laser. In the absence of an atom we measure the reflection of a diagonally polarized probe field as a function of laser frequency $\nu$. The output state of the interferometer is $|\psi_{out}\rangle = (1/\sqrt{2})(|r_V| e^{i \phi_0(\nu)}|V\rangle + r_c(0)|H\rangle)$ where $\phi_0(\nu)=\nu/\nu_{FSR}$ is the relative phase accumulated between the two arms, $\nu_{FSR}$ is the interferometer free spectral range, $ r_c(0)=|r_c(\nu)|e^{i\phi_c(\nu)}$ and $|r_c(\nu)|$ and $\phi_c(\nu)$ the reflection coefficient and phase of the empty cavity respectively. We measure the power in the $D$ and $A$ ports as a function of probe detuning and analyze the sum and difference of the two detectors: 
\begin{align}
\mathcal{D}+\mathcal{A}&\sim\frac{1}{2}\left(|r_V|^2+|r_c(0)|^2\right)\,\label{eq:DpA}\\
\mathcal{D}-\mathcal{A}&\sim\mathrm{Re}\left(|r_V| r_c(0) e^{-i \phi_0(\lambda)}\right)\label{eq:DmA}
\end{align}
Figure \ref{fig:interfFSR} shows a measurement of both the sum and difference of the output intensities at the $A$ and $D$ interferometer ports. The sum shows a dip in reflected power at the cavity resonance, which arises from the finite losses of the cavity. In the differential signal the resonance is visible as a $2\pi$ phase slip in the interferometer signal across the cavity resonance. The red lines are fits of Eq. \ref{eq:DpA} and \ref{eq:DmA} with $\nu_{FSR}$, $\kappa_{wg}$, $\kappa_{sc}$ and a global phase and amplitude as free parameters. We obtain $\nu_{FSR}=33\,\mathrm{GHz}$, $\kappa_{wg}=20.3\,\mathrm{GHz}$ and $\kappa_{sc}=5.2\,\mathrm{GHz}$, yielding $k=\kappa_{wg}/\kappa=0.8$, where $\kappa=\kappa_{sc}+\kappa_{wg}$.

\subsection{Cavity tuning}
At room temperature the cavity resonance is at $779.8\,\mathrm{nm}$, selected to be slightly to the blue of the atomic $^{87}$Rb resonance at $780.03\,\mathrm{nm}$. We thermally tune the cavity resonance by applying a 802 nm laser beam focused to a $2.1\,\mathrm{\mu m}$ beam waist on the absorptive silicon patch shown in Figure 1b. Heating the cavity allows for thermal tuning of the cavity resonance to the red with a measured tuning coefficient of $\sim 0.4\,\mathrm{nm/mW}$. We have observed reversible tuning up to $+2\,\mathrm{nm}$ from the room-temperature resonance. 

To lock the cavity on the atomic resonance we use a similar scheme as we use to lock the interferometer. We measure the reflection of the same $780\,\mathrm{nm}$ probe beam and lock the cavity resonance by applying feedback to the heater laser intensity. We dither the heater-intensity and perform a lock in measurement and feedback using the FPGA.
The feedback is applied during the Raman cooling sequence, when the probe beam does not disturb the atom. The cavity is typically locked $\sim 1-2$ linewidths to the blue of the $^{87}$Rb resonance to account for a small additional heating of the cavity by the optical tweezer. Pointing alignment of the heater laser on the PC is periodically optimized by scanning the heater laser position using a scanning piezo mirror minimizing the heater laser intensity required for locking the cavity.

\subsection{Measurement sequence for switching experiment}
\begin{figure}[tpb]
\begin{center}
\includegraphics[width=0.8 \textwidth]{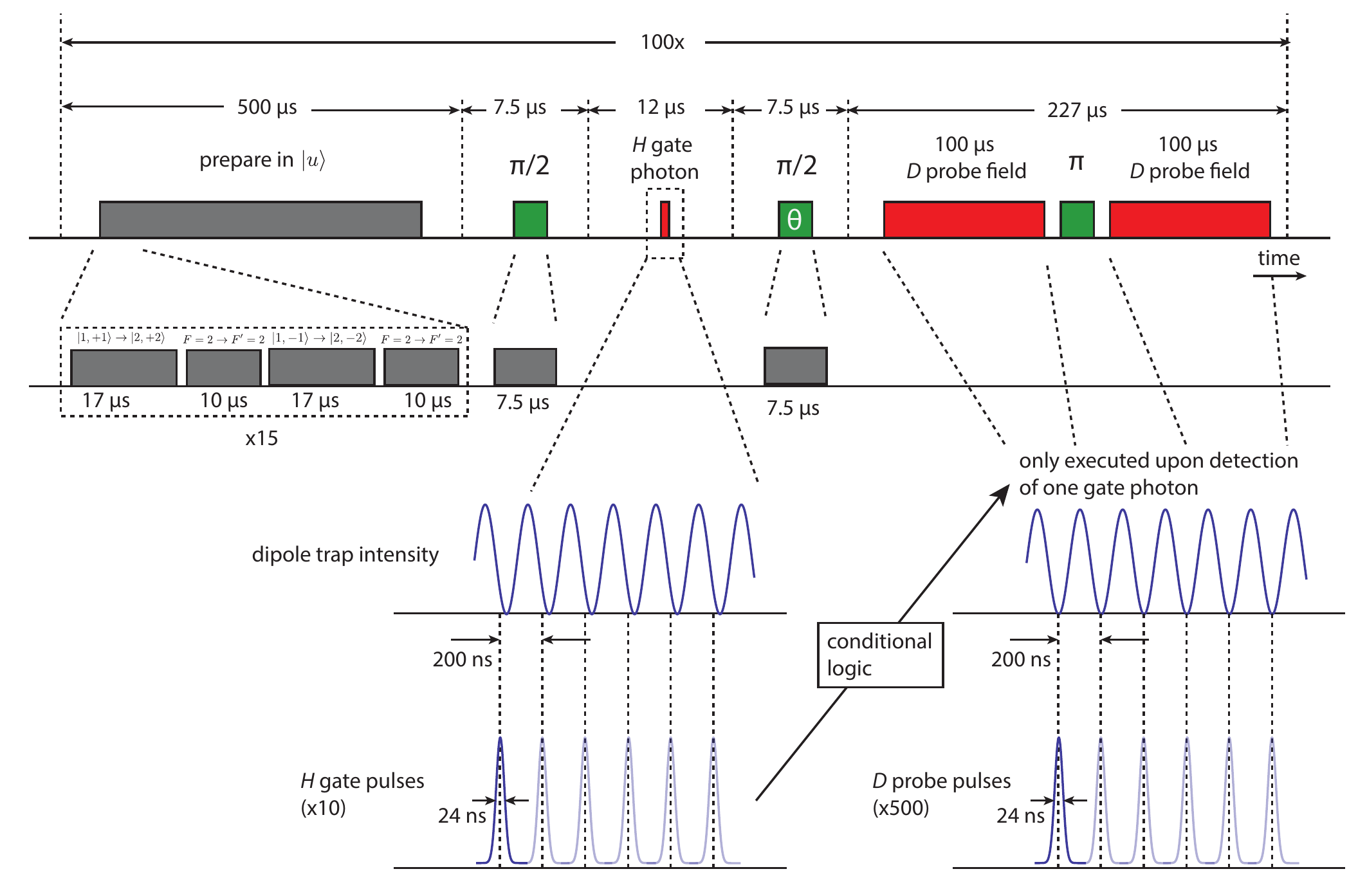}
\caption{\label{fig:seq} A schematic representation of the switch sequence. See text for details.}
\end{center}
\end{figure}

Figure \ref{fig:seq} shows a detailed version of the sequence used for the switch experiment. The first $500\,\mathrm{\mu s}$ consist of preparing the atom in the $|u\rangle$ state by means of microwave transfer and optical pumping (see section \ref{sect:uwpump}). Following, the atom is put in the superposition $|u\rangle +|c\rangle$ by means of a $7.5\,\mathrm{\mu s}$ long $\pi/2$ microwave pulse. Then 10 Gaussian (FWHM of $24\,\mathrm{ns}$) $H$-polarized gate pulses are applied at instances that the dipole trap is at its minimal intensity. A second $\pi/2$ microwave pulse with variable phase $\theta$ rotates the atomic superposition to its final state. The atomic state is detected with 500 Gaussian probe pulses at times that the dipole trap intensity is minimal, followed by a $15\,\mathrm{\mu s}$ long $\pi$ pulse and a second identical readout sequence. For the data with the gate pulse a fast conditional logic circuit (Lattice ispMACH LC4256ZE) prohibits the execution of the readout sequence if no photon was detected during the a several $\mu$s-wide window around the gate pulses. This prevents unnecessary heating of the atom by the readout at instances that no gate photon was detected. The complete sequence is  typically repeated 100 times for one atom. 

For the measurements of the photon phase presented in Figure 2 we use a similar method of pulsed probing using the dipole trap modulation without applying the microwave modulation and the gate pulse. For the data in Figure 3, no dipole trap modulation was applied. 

\section{Experimental methods}

\subsection{Internal state preparation}\label{sect:uwpump}
For the switching experiments we perform state preparation into the $|F=1,m_F=0\rangle$ magnetic sub-level. Conventional methods for state preparation in magnetic sub-levels involve optical pumping with well defined polarization such that one internal state is dark to the optical pumping process. However, in the vicinity of the PC obtaining a clean polarization is challenging because of unavoidable scattered light. To achieve efficient optical pumping while being trapped near the PC we employ a combination of optical pumping and coherent microwave transfer. We apply light resonant with the $F=2\rightarrow F'=2$ transition of the D2-line to deplete the $F=2$ manifold. Simultaneously, we perform a coherent microwave transfer between the $|F=1,m_F=\pm1\rangle$ and $|F=2,m_F=\pm2\rangle$ sub-levels. As a result the atomic population accumulates in the dark $|F=1,m_F=0\rangle$ sub-level. We toggle the microwave pulses (each $17\,\mathrm{\mu s}$) and the optical pulses ($10\,\mathrm{\mu s}$) sequentially but have observed similar behavior with the optical beam continuously on and $25\,\mathrm{\mu s}$ microwave pulses. The optical intensity is chosen to have similar optical pumping rates and microwave transfer rates. We use the stretched $|F=1,m_F=\pm1\rangle \rightarrow |F=2,m_F=\pm2\rangle$ transitions instead of $|F=1,m_F=\pm1\rangle \rightarrow |F=2,m_F=\pm1\rangle$ because of the larger Clebsch Gordan coefficients and slightly $\sigma$-polarized nature of our microwave field, leading to a faster pumping rate. Under conditions with $25\,\mathrm{\mu s}$ microwave pulses and continuous optical pumping we find that after $5$ cycles the atom is with $\sim 90\%$ probability in the $|F=1,m_F=0\rangle$ state, and the pumping rate is well fit by an exponential time-constant of $\tau_{1/e}=57 \,\mathrm{\mu s}$.

\subsection{Single shot readout and verification of the atom presence}
A single readout measurement of the atom consists typically of 500 $D$-polarized probe pulses (see Figure \ref{fig:seq}) over which we detect $\bar n_A^1\simeq 6.2$ and $\bar n_A^0\simeq 0.2$ for the cases with and without an atom present in the $F=2$ manifold (see Figure 4b). We observe no distinction between an atom present in the uncoupled $F=1$ manifold and no atom present, confirming that the $F=1$ state is not coupled to the cavity field. 
We assign events with $n>1$ to have an atom present in the $F=2$ manifold. For this threshold the measured fidelity for correctly assigning the cases without and with an atom in the $F=2$ manifold is 97.2\% and 93.7\% respectively, yielding a combined readout fidelity of 95\%. Poissonian distributions with the measured mean photon numbers would yield a readout fidelity of 98\%. The reduction from the expected value is mostly due to an increased probability of events with low photon numbers, which we attribute to a finite optical pumping probability out of $F=2$ during the readout period. 

\begin{figure}[tpb]
\begin{center}
\includegraphics[width=0.5 \textwidth]{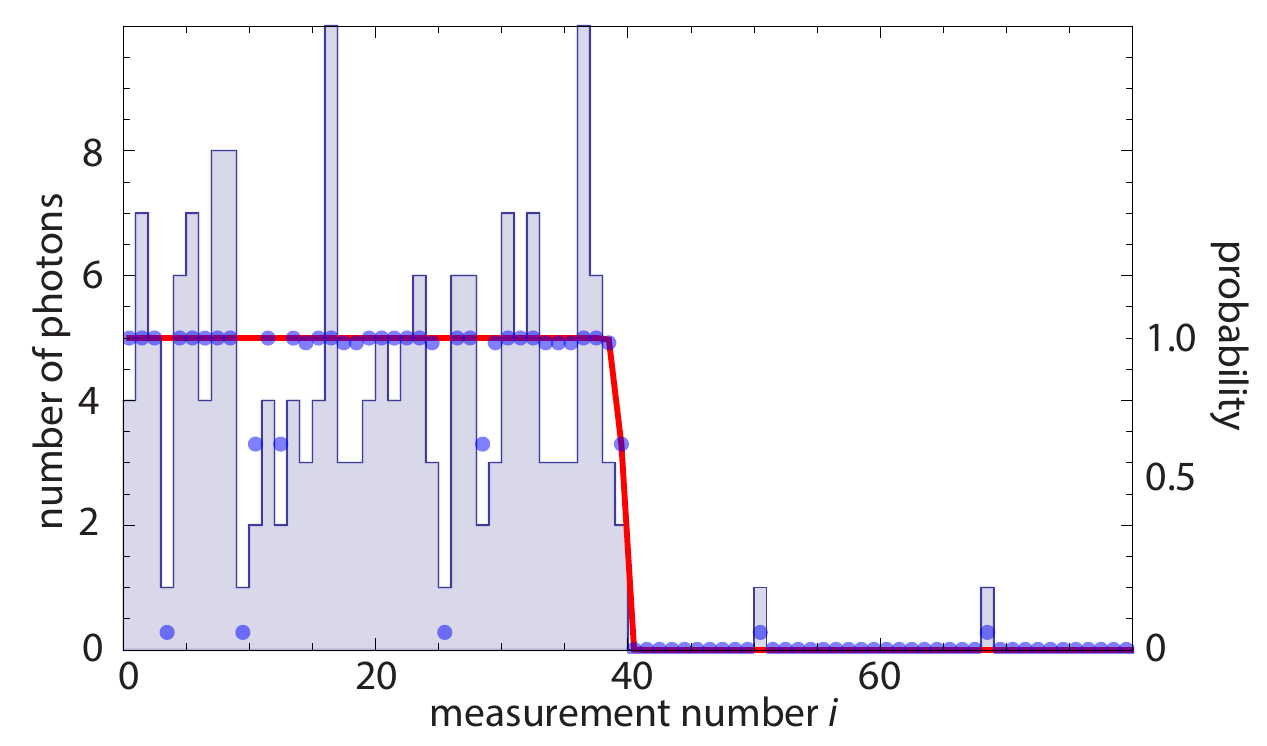}
\caption{\label{fig:postsel}Determining the presence of a single atom. 80 readout measurements (each consists of 500 $24\,\mathrm{ns}$ pulses) are performed on the same atom. The histogram shows the number of photons counted per readout measurement. The blue dots show the probability $P_i$ for a single measurement that an atom was present and the red line $P^C_i$ that an atom was present at the $i$-th measurement based on all measurements in the dataset (see text).}
\end{center}
\end{figure}

We typically repeat a measurement $100$ times per trapped atom. After this period the atom is lost from the trap with high probability. Hence, we analyze our data to select only events where an atom was present or not with high certainty. 
Based on only the collected photons for a single measurement $i$ and the Poissonian distributions above, we determine the individual probability $P_i$ that an atom was present during that measurement. However, by combining the 100 individual probabilities $P_i$ of all measurements and assuming that the atom escapes the trap once and for all at a certain point in time, we obtain a more accurate probability $P^C_i$ of having had an atom at the $i$-th measurement. 

A typical dataset of 80 measurements on the same atom is shown in Figure \ref{fig:postsel}. The histogram shows the number of counted photons for each measurement, the blue dots $P_i$ and the red line $P^C_i$. We typically condition on a probability for the atom to be present of $99.9\%$. For this particular case it implies the atom is lost at the $38$-th pulse.

For the measurements of the atomic spin state the uncoupled state $|u\rangle$ is indistinguishable from having no atom present in the trap. Therefore, for a typical experiment every $4$th measurement we perform a control measurement by optically pumping the atom in the $F=2$ manifold and verify its presence. In case the atom presence is not confirmed we discard all data after the last control measurement where an atom was detected.

For the switching experiments we perform two readout sequences (see Figure \ref{fig:seq}) to ensure the atom was in the $\{|u\rangle,|c\rangle\}$ subspace. During the first readout sequence the atom is projected on either $|u\rangle$ or $|c\rangle$ by the first scattered probe photon and the following scattering events distribute $|c\rangle$ over the $F=2$ manifold but do not affect $|u\rangle$. Subsequently, we apply a microwave $\pi$-pulse transferring $|u\rangle$ to $|c\rangle$ followed by a second readout pulse. If in either of the two readout sequences the atom is detected then the atom was in the $\{|u\rangle,|c\rangle\}$ subspace. 

In the data in Figure 2a we can verify the presence of the atom only for certain values of $\phi_V$ ($-0.3\leq\phi_V\leq0.2$ and $0.8\leq\phi_V\leq1.2$), where the mean photon numbers in $A$ and $D$ are significantly affected by the presence of an atom. For these values of $\phi_V$ we determine that in $89\%$ of the cases an atom was present during the first 10 measurements on each atom. We apply a correction of $11\%$ to the other data in Figure 2 obtained at $\phi_V$ where the contrast was not sufficient to verify the presence of the atom. 

\section{Supplementary discussion}
\label{sect:suppdisc}
In this section we discuss several aspects of the experiments presented in the main text. 

The proximity of a surface can alter the spontaneous emission rate of an atom substantially even in the absence of resonant structures \cite{klimov04}. 
We estimate the change in spontaneous emission due to the proximity of the SiN waveguide by performing Finite-Difference-Time-Domain (FDTD) simulations of a dipole located at $180\,\mathrm{nm}$ from the surface of a waveguide with equal dimensions to our PC. We perform simulations for an unpatterned waveguide and for a waveguide with holes spaced at $315\,\mathrm{nm}$, corresponding to the  mirror sections of the PC but no resonant cavity structure. We find that the enhancement of the decay rate for all dipole orientations is between $1$ and $1.25$, hence only a perturbation compared to the cavity enhanced decay rate. In the rest of our analysis we neglect this effect.We expect non-radiative contributions to the decay of the atomic excited state to be negligible because the imaginary part of the dielectric permittivity of SiN in this wavelength range is small. Additionally, non-radiative decay processes would be represented in the data as a frequency independent decay enhancement, which we do not observe (see Figure 1d).

In Figure 1d, we measure the lifetime of the atomic excited state. Given a possible shot-to-shot variation in cooperativity (resulting \emph{e.g.} from atomic motion), fitting a single exponential decay to the data gives a conservative estimate of the fastest decay rate in the ensemble. The data is accumulated over a window that begins $1\,\mathrm{ns}$ after the end of the excitation pulse, to ensure that background light from the falling edge of the pulse is excluded. However, this has the effect of systematically biasing the measurement away from fast decay rates. Therefore, the cooperativity of $\eta=7.7$ measured from the decay rate should be interpreted as a conservative estimate of the maximum cooperativity in an ensemble. We also attribute the linewidth in the inset of Fig. 1d to this effect: the measured linewidth is 43 GHz, while the independently recorded cavity linewidth is 25 GHz (Fig. \ref{fig:interfFSR}). 

For the measurements presented in Figure 2 various experimental imperfections contribute to reduction of the fringe visibility $(\mathcal{P}_{max}-\mathcal{P}_{min})/(\mathcal{P}_{max}+\mathcal{P}_{min})$. Imperfect balancing of the interferometer accounts for 5\% reduction of the visibility, which we extract from the measurements without an atom. Additional reduction can arise from finite saturation. To estimate the influence of saturation we have performed a measurement similar to Figure 2a at $\phi_\mathrm{V}=\pi$ and an $8$ times smaller driving intensity. We observe a maximum signal in the $A$ port of $\mathcal{A}_1/\mathcal{P}_1=85\%$, compared to $75\%$ as shown in Figure 2. 

In our experiment the magnetic field axis is aligned orthogonal to the linearly polarized cavity field. Therefore, linearly polarized photons emitted into the cavity mode can leave the atom in a final state different from the initial state and reveal which-path information in the interferometer and therefore reduce the fringe visibility. We estimate this contribution to be $\sim10\%$ of the scattering events. This effect can be suppressed by aligning the magnetic field axis and the cavity field polarization. The same effects are present in the measurements presented in Figure 4 where such scattering events move the atom out of the $\{|u\rangle,|c\rangle\}$ subspace.

Additional reduction of the fringe visbility could arise from positioning uncertainty of the atom with respect to the cavity mode (\emph{e.g.} from thermal motion of the atom in the dipole trap) that gives rise to a fluctuating $\eta$. The cavity mode is a standing wave along the cavity axis and the effect of position uncertainty on our measurements depends strongly on the precise distribution of positions. Assuming complete uncertainty in the position along the cavity axis we estimate that this does not account for more than $20\%$ in reduction of the fringe visibility.

In Figure 4 we typically route $\bar n_A=6.2$ photons which is optimized to have a high readout fidelity and minimal heating of the atom, thereby increasing the number of repetitions of the experiment with the same atom. In the same configuration we have routed up to $\bar n_A =14$ photons per readout pulse and in an unmodulated trap we have routed up to $\bar n_A \simeq 75$ detected photons after which the atom is optically pumped to the $F=1$ manifold with $\sim50\%$ probability.

In Figure 4b we present two datasets with an applied gate field: one where the switch state is conditioned ($P_{on}^1$) on having detected at least one gate photon, and one where it is not conditioned ($P_{on}^{uc}$) . The readout of the switch state is triggered by the arrival of a gate photon in a broad time window. The conditioned and unconditioned datasets are extracted from the same measurement by dividing the events based on the arrival time of the gate photon. If the gate photon arrived during one of the 24 ns pulses, then the measurement was included in the conditioned dataset. Otherwise, the photon is assumed to be a background event uncorrelated with the gate field, and the measurement is included in the unconditioned dataset. Careful analysis of the arrival times shows that about 70\% of the background events are fluorescence from the dipole trap, while 30\% are actually leaked gate photons that arrive at the wrong time. Since most of these photons arrive at times when the dipole trap intensity is high, they are mostly detuned by more than $\eta \gamma$, on average, and can be safely approximated as background events. 

Finally, in Figure 4c an additional reduction of the fidelity arises from a combination of imperfect internal state preparation and readout, dephasing and microwave-pulse accuracy. All these effects are present in the data without a gate field (red curve) and amount to a maximum reduction of the fidelity of $\sim10\%$.

\section{Theoretical analysis}
\label{sect:theory}
In this section, we outline the theoretical framework used to analyze our experimental observations. 

\subsection{Theoretical model}

\begin{figure}[htbp]
\begin{center}
\includegraphics{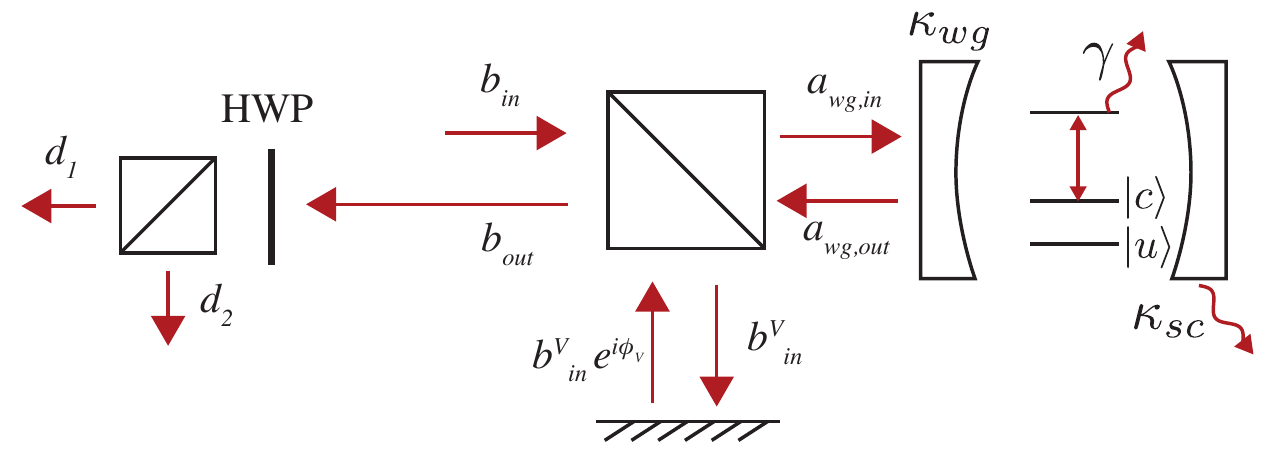}
\caption{\label{fig:modes} Modes at various points in the interferometer, see text for details.}
\end{center}
\end{figure}

We consider an atom interacting with a single mode, single sided optical cavity. In a frame rotating with the incident laser, the Hamiltonian governing the atom-cavity dynamics is:
\begin{equation}
\label{eq:hamiltonian}
H_\mathrm{ac} = \frac{1}{2}\Delta_a \sigma_z + \Delta_c a^\dag a + g(a^\dag \sigma + a \sigma^\dag), 
\end{equation}
where   $\sigma$ and $a$ are the atomic and photonic lowering operators, $\sigma_z$ is the atomic pseudo spin operator, $\Delta_a = \omega_a - \omega_L$ and $\Delta_c = \omega_c - \omega_L$ are the detunings between the bare atomic ($\omega_a$), cavity ($\omega_c$) and laser ($\omega_L$) frequencies, and $2g$ is the single-photon Rabi frequency. Note that $\Delta_a = -\delta$ in the main text. In the presence of atomic excited state decay $(\gamma$) and cavity decay into the waveguide $(\kappa_{wg})$ and into other dissipation channels $(\kappa_{sc})$ the quantum dynamics is governed  by Heisenberg-Langevin equations of motion \cite{carmichaelBook1}:
\begin{align}
\label{HLa}
\dot{a}(t) &= -i g \sigma(t) - (\kappa/2 + i \Delta_c) a(t) - \sqrt{\kappa_{wg}} a_{wg,in}(t) - \sqrt{\kappa_{sc}} \xi_{sc}(t) \\
\label{HLsigma}
\dot{\sigma}(t) &= i g a(t) \sigma_z(t) - (\gamma/2 + i \Delta_a) \sigma(t) + \sqrt{\gamma} \xi_{at}(t) \sigma_z(t) \\
\label{HLsigmaz}
\dot{\sigma_z}(t) &= -2 i g (\sigma^\dag(t) a(t) - \sigma(t) a^\dag(t)) - \gamma (\sigma_z(t)+1) - 2 \sqrt{\gamma} \xi^\dag_{at}(t) \sigma(t) - 2  \sqrt{\gamma} \sigma^\dag(t) \xi_{at}(t)
\end{align}
where $a_{wg,in}(t)$ is the input field operator representing the cavity-waveguide coupling and $\xi_{sc}(t)$ and $\xi_{at}(t)$ are noise operators corresponding to other cavity dissipation channels and atomic spontaneous emission into other modes, respectively. The cavity output field is described by the input-output relation:
\begin{equation}
\label{inputoutputwg}
a_{wg,out}(t) - a_{wg,in}(t) = \sqrt{\kappa_{wg}} a(t)
\end{equation}
Following the notations defined in Fig. \ref{fig:modes}, the interferometer input and output fields $\vec{b}_{in}$, $\vec{b}_{out}$ are:
\begin{align}
\vec{b}_{in} &= b_{in}^H \hat{\epsilon}_h + b_{in}^V \hat{\epsilon}_v \\
\vec{b}_{out} &= (b_{in}^H + \sqrt{\kappa_{wg}} a) \hat{\epsilon}_h + b_{in}^V e^{i \phi_V} \hat{\epsilon}_v
\end{align}
where $\hat{\epsilon}_{h/v}$ are unit vectors denoting horizontal and vertical polarization. Here, we have treated the action of the $V$-polarized reference arm as perfect reflection with a phase shift $\phi_V$. 
Using a HWP to orient the detection basis at an angle $\theta'$ with respect to the $H$ axis, the fields at the two detectors are given by:
\begin{align}
\label{d1}
d_1 &= \vec{b}_{out} \cdot (\cos \theta' \hat{\epsilon}_h + \sin \theta' \hat{\epsilon}_v) \\
\label{d2}
d_2 &= \vec{b}_{out} \cdot (-\sin \theta' \hat{\epsilon}_h + \cos \theta' \hat{\epsilon}_v)
\end{align}
If the input field is linearly polarized at an angle $\theta$ with respect to the $H$ axis, then we can define two input modes $b_{s}$ and $b_{v}$ such that $\vec{b}_{s} = b_{s} \left( \cos \theta \hat{\epsilon}_h + \sin \theta \hat{\epsilon}_v \right)$ and $\vec{b}_{v} = b_{v} \left(-\sin \theta \hat{\epsilon_h} + \cos \theta \hat{\epsilon_v}\right)$. The mode $b_v$ is orthogonal to the input field and is not driven.

In practice, we fix $\theta^\prime=\pi/4$ and adjust $\theta$ to compensate for the effects of cavity losses and finite reflectivity in the $V$-polarized arm of the interferometer, as discussed in the next section. In this case, the output modes are given by:
\begin{eqnarray}
\label{eq:d1inout}
d_1 &=& A_1 b_s + a \sqrt{\frac{\kappa_{wg}}{2}} + C_1 b_v \\
\label{eq:d2inout}
d_2 &=& A_2 b_s - a \sqrt{\frac{\kappa_{wg}}{2}} + C_2 b_v
\end{eqnarray}
with $A_1 = (\cos \theta + e^{i \phi_V} \sin \theta)/\sqrt{2}$, $A_2 = (-\cos \theta + e^{i \phi_V} \sin \theta)/\sqrt{2}$, $C_1 = (e^{i \phi_V}\cos \theta - \sin \theta)/\sqrt{2}$ and $C_2 = (e^{i \phi_V}\cos \theta - \sin \theta)/\sqrt{2}$. Note that for  $\theta=\pi/4$ and $\phi_V=0$ these equations are identical to the input-output relations for a symmetric cavity driven from one side by $b_s$ and from the other side by $b_v$.

\subsection{Linear response}
\label{linresponse}
When the driving field is weak, the atom is nearly always in its ground state and we can approximate the action of the operator $a \sigma_z$ with $-a$. In this case, the expectation values of Eq. (\ref{HLa}-\ref{HLsigma}) form a closed system of differential equations that we can solve exactly to find the response to a slowly-varying incident coherent field $\langle a_{wg,in}\rangle$: 
\begin{eqnarray}
\<a\> &=& \frac{-\sqrt{\kappa_{wg}}}{\tilde{\kappa}} \frac{\<a_{wg,in}\>}{1+\frac{g^2}{\tilde{\kappa}\tilde{\gamma}}} \\
\<\sigma\> &=& \frac{i g \sqrt{\kappa_{wg}}}{\tilde{\kappa}\tilde{\gamma}}\frac{\<a_{wg,in}\>}{1+\frac{g^2}{\tilde{\kappa}\tilde{\gamma}}}
\end{eqnarray}
where we have introduced complex decay rates $\tilde{\kappa} = \frac{\kappa}{2} + i \Delta_c$ and $\tilde{\gamma} = \frac{\gamma}{2} + i \Delta_a$, and defined the cooperativity $\tilde{\eta} = g^2 / \tilde{\kappa} \tilde{\gamma}$. On resonance, it reduces to $\eta = 4 g^2/\kappa \gamma$. The reflected field from the cavity is given by Eq. \ref{inputoutputwg}:
\begin{eqnarray}
\label{refl}
\<a_{wg,out}\> = \<a_{wg,in}\> \frac{\tilde{\kappa}(1+\tilde{\eta}) - \kappa_{wg}}{\tilde{\kappa}(1+\tilde{\eta})}  \rightarrow \<a_{wg,in}\>\frac{\eta - 1}{\eta + 1},
\end{eqnarray}
where the final limit is taken on resonance and with $\kappa_{wg} = \kappa$. This expression captures the key result of the atom-photon interaction: the reflection coefficient of the cavity changes from $-1$ to $+1$ in the presence of a strongly coupled atom with $\eta > 1$.

The output fields at the two ports of the interferometer are given by
Eq. (\ref{eq:d1inout}-\ref{eq:d2inout}). For the data presented in the paper, the angle $\theta$ was chosen to null the light in one port ($d_1$) in the absence of an atom, with the probe field on resonance with the cavity. This is accomplished when $\tan \theta = (2 k - 1)$, with $k=\kappa_{wg}/\kappa$. For this choice of $\theta$ the $d_1$ and $d_2$ ports correspond to the $A$ and $D$ ports in the main text respectively. The output field is given by:
\begin{eqnarray}
\<d_1\> = \frac{\<b_s\>}{2 \sqrt{1+2k(k-1)}} \frac{(e^{i \phi_V}(2k-1)+1)(1+\tilde{\eta})- \kappa_{wg}/\tilde{\kappa}}{1+\tilde{\eta}}  \rightarrow \<b_s\>\frac{\eta}{1+\eta}\label{eq:d1}\\
\<d_2\> = \frac{\<b_s\>}{2 \sqrt{1+2k(k-1)}} \frac{(e^{i \phi_V}(2k-1)-1)(1+\tilde{\eta}) + \kappa_{wg}/\tilde{\kappa}}{1+\tilde{\eta}} \rightarrow \<b_s\>\frac{1}{1+\eta}\label{eq:d2}, 
\end{eqnarray}
where the final expression is evaluated on resonance with $k = 1$ and $\phi_V=0$. In this case, the fields at the ports $d_1$ and $d_2$ are identical to the reflection and transmission outputs of a symmetric cavity with an atom. Within this linear approximation, the  
intensity at the output ports is given by  $\<d_1^\dag d_1\> = |\<d_1\>|^2$. 

\begin{figure}[tpb]
\begin{center}
\includegraphics[width=0.5 \textwidth]{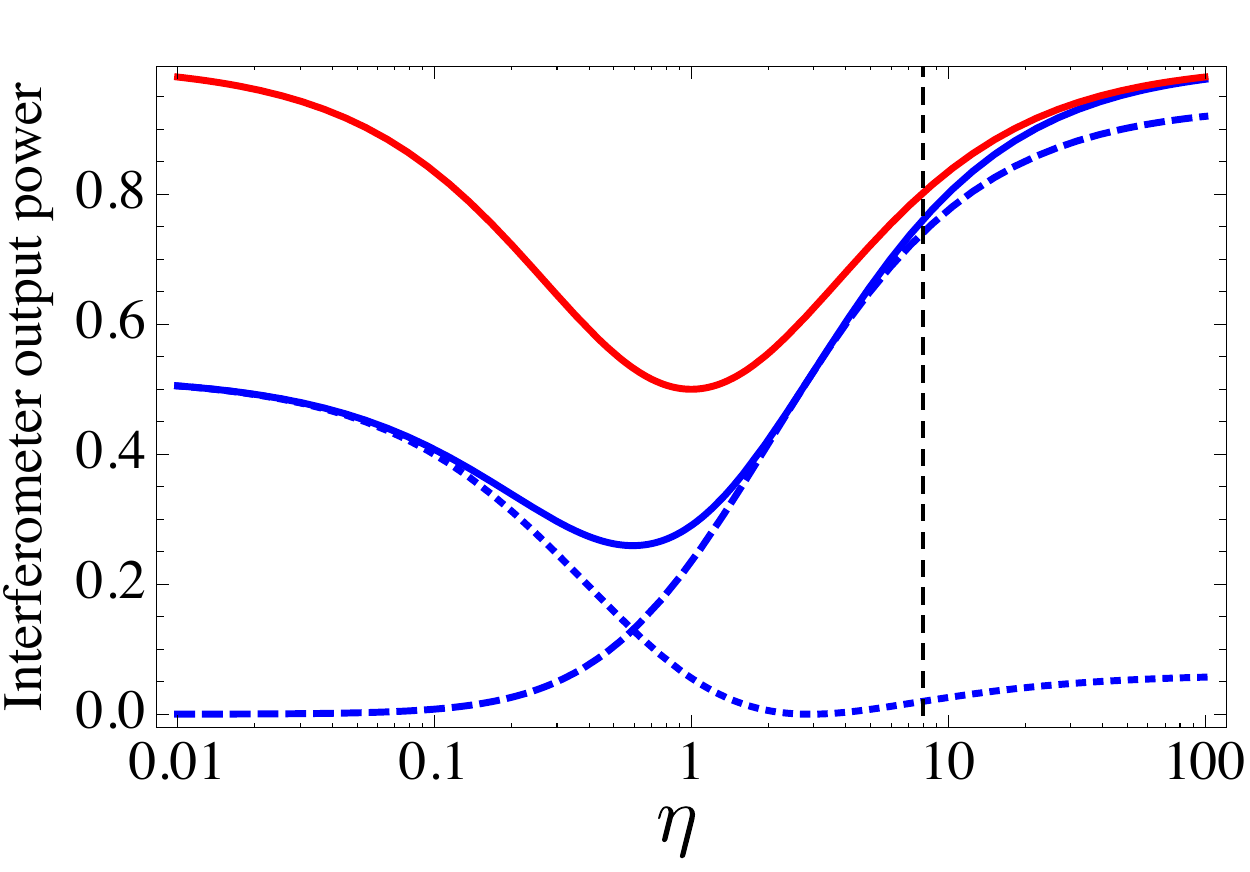}
\caption{\label{fig:interfPower}Interferometer output as a function of cooperativity given by Eq. \ref{eq:d1} and \ref{eq:d2}. The red solid curve shows the total power in the absence of cavity losses, blue dashed, dotted and solid lines show the power $|\<d_1\>|^2$, $|\<d_2\>|^2$ and $|\<d_1\>|^2 +|\<d_2\>|^2$ respectively including cavity losses of $k=0.8$. }
\end{center}
\end{figure}

Figure \ref{fig:interfPower} shows the output power of the interferometer on resonance ($\Delta_a=\Delta_c=0$) versus cooperativity, both without losses (red, $k=1$) and with losses for our cavity parameters (blue, $k=0.8$).  At $\eta=1/k$ the cavity reflection vanishes, and the interferometer reflection is $0.5$. In the absence of an atom ($\eta=0$) the cavity reflection is determined by the cavity losses. For an atom strongly coupled to the cavity ($\eta\gg1$) the light is blocked from entering the cavity, therefore the cavity losses are suppressed.  For our parameters ($\eta\simeq8$) the reflection on resonance without an atom is $ |\<d_1\>|^2+ |\<d_2\>|^2=0.35$ and is expected to increase by $(|\<d_1\>|^2+|\<d_2\>|^2)_{\eta=8}/(|\<d_1\>|^2+|\<d_2\>|^2)_{\eta=0}=1.46$ due to the presence of an atom. The expected maximum fraction of the power switched from $D$ to $A$ is $|\<d_1\>|^2/(|\<d_1\>|^2+|\<d_2\>|^2)=0.97$. 

\subsection{Saturation behavior}

The treatment above made use of the approximation $\<\sigma_z\> \approx -1$. This is valid for input field amplitudes such that the excited state population $\sim \langle \sigma^\dag \sigma \rangle$ remains small:
\begin{equation}
\<\sigma^\dag \sigma\> \approx |\<\sigma\>|^2 = \frac{\kappa_{wg}}{\left|\tilde{\kappa}\tilde{\gamma}\right|^2} \frac{g^2}{\left|1+\tilde{\eta}\right|^2} |\<a_{wg,in}\>|^2 \ll 1.
\end{equation}
In the limit of large $\eta$ and neglecting other cavity loss channels, this corresponds to $|\<a_{wg,in}\>|^2 \ll \eta \gamma/2$. That is, the rate of incident photons should be much less than one per excited state lifetime.

For stronger input fields, this approximation breaks down due to atomic saturation. This has two consequences. First, we can no longer make the approximation $\<\sigma_z\> \approx -1$. Second, we can no longer factor expectation values of operator products. To describe saturation dynamics of our system, we make use of the hierarchy of our experimental parameters ($\kappa \gg (g, \gamma)$), which  allows us to integrate out the cavity degree of freedom. Following \cite{rice88}, in the limit of large $\kappa$ we find:
\begin{equation}
\label{aint}
 a(t) = \left[- i g \sigma(t) - \sqrt{\kappa_{wg}} a_{wg,in}(t) - \sqrt{\kappa_{sc}} \xi_{sc}(t)\right]/\tilde{\kappa}
\end{equation}
which yields the following atomic dynamics (assuming vacuum for the $\xi_{sc}$ and $\xi_{at}$ operators):
\begin{eqnarray}
\dot{\<\sigma\>} &=& -\left(\tilde{\gamma} + \frac{g^2}{\tilde{\kappa}}\right) \<\sigma\> - \frac{i g}{\tilde{\kappa}} \sqrt{\kappa_{wg}} \<a_{wg,in}\> \<\sigma_z\> \\
\dot{\<\sigma_z\>} &=& -\left(\gamma + \frac{g^2 \kappa}{|\tilde{\kappa}|^2}\right) (\<\sigma_z\> + 1) + 2 i g \sqrt{\kappa_{wg}}\left(\frac{\<\sigma^\dag\> \<a_{wg,in}\>}{\tilde{\kappa}} - \frac{\<\sigma\> \<a_{wg,in}^\dag\>}{\tilde{\kappa}^*}\right).
\end{eqnarray}
Focusing on the resonant cw case ($\Delta_a = \Delta_c = 0$) and introducing a dimensionless amplitude of the driving field $Y = \frac{4 g}{\kappa \gamma} \sqrt{\kappa_{wg}} \<a_{wg,in}\>$, we find the following steady state solution:
\begin{eqnarray}
\<\sigma\> &=& \frac{i Y (1+\eta)}{2 Y^2 + (1+\eta)^2} \\
\<\sigma_z\> &=& \frac{-(1+\eta)^2}{2 Y^2 + (1+\eta)^2}.
\end{eqnarray}

Choosing $\tan \theta = 2 k -1$ and $\phi_V=0$ again, we find for the output ports of the interferometer:
\begin{eqnarray}
\<d_1^\dag d_1\> &=& \frac{\gamma}{2 \eta} k Y^2 \frac{\eta^2}{2 Y^2 + (1+\eta)^2}\label{eq:pow1}\\
\<d_2^\dag d_2\> &=& \frac{\gamma}{2 \eta} k Y^2 \frac{(2Y^2+1)(1-2k)^2+2(1-k)(1-2k)\eta+(1-k)^2\eta^2}{k^2(2 Y^2 + (1+ \eta)^2)}.\label{eq:pow2}  
\end{eqnarray}
The dimensionless driving intensity is $Y^2 = \frac{4 \eta}{\gamma} k |b_s|^2 \cos^2 \theta$, in terms of the input field intensity $|b_s|^2$ (which has units of photons/second). 

The two-photon correlation functions are calculated using:
\begin{align}
g_i^2(\tau) = \frac{\<d_i^\dag(t)d_i^\dag(t+\tau)d_i(t+\tau) d_i(t) \>}{\<d_i^\dag(t) d_i(t)\>^2}\label{eq:g2}
\end{align}
with the operator $d_1$ for the $A$ port, and $d_2$ for the $D$ port. The time dependence has been worked out analytically for the case of a double-sided cavity in the bad cavity limit in Ref. \cite{carmichaelBook1}. Here, we calculate $g^2(\tau)$ numerically from the master equation using the Hamiltonian (\ref{eq:hamiltonian}) together with the quantum regression theorem \cite{carmichaelBook1}.

The data in Figure 3 was taken in the presence of the dipole trap which imposes a position-dependent AC-Stark shift, which fluctuates because of the finite kinetic energy of the atom. To account for this effect, we average equations (\ref{eq:pow1}$-$\ref{eq:g2}) over a Gaussian distribution of atomic detunings with a standard deviation of $60\,\mathrm{MHz}$. The resulting averaged intensities and correlation functions are shown as the solid lines in Figure 3. We estimate the maximum differential light shift between the ground and excited state in our trap to be $\sim130\,\mathrm{MHz}$.
In the normalization of the averaged $g^2(\tau)$ we account for the detuning-dependent intensity at the detector, with the assumption that the timescale on which the detuning fluctuates is much faster than the window over which photon data is accumulated (tens of $\mathrm{ms}$) but slower than the excited state dynamics in Figures 3b-c. 

\subsection{Analysis of the quantum phase switch}

In this section we consider the effect of imperfections such as photon loss and multiphoton excitations on the operation of the quantum phase switch. 

Photon loss is included in the Heisenberg-Langevin equations (\ref{HLa} - \ref{HLsigmaz}) in the form of coupling to three photonic modes: the waveguide mode $a_{wg}$, a second cavity loss mode $\xi_{sc}$, and a spontaneous emission mode $\xi_{at}$. For low input intensities, with less than one photon per bandwidth,  the optical response is linear. In this case  the combined evolution of the atomic and photonic systems can be represented as:
\begin{equation}
U = \sum_{i,j} \left(r_{ij,c} \ket{c}\bra{c} + r_{ij,u} \ket{u}\bra{u}\right) a_i^\dag a_j, 
\end{equation}
where $r_{ij,c}$ and $r_{ij,u}$ denote the scattering amplitudes coupling modes $i,j$, with an atom in the coupled or uncoupled states. $a_i$ corresponds to one of the modes $\{a_{wg},\xi_{sc},\xi_{at}\}$. The scattering amplitudes can be found by using input-output relations similar to Eq. (\ref{inputoutputwg}) for the modes $\xi_{sc}$ and $\xi_{wg}$, together with the linear response as described in \ref{linresponse}. Since only the mode $a_{wg}$ is driven, the important terms are the coefficients for $a_{wg}^\dag a_{wg}$, $\xi_{sc}^\dag a_{wg}$ and $\xi_{at}^\dag a_{wg}$, which we denote as $r_i$, $t_i$ and $l_i$, respectively, with the appropriate atomic state subscript. They obey $|r_i|^2 + |t_i|^2 + |l_i|^2 = 1$. These are related to microscopic parameters of the system as follows: 
\begin{align}
\label{ru}
r_u &= 1-\frac{\kappa_{wg}}{\tilde{\kappa}} \\
\label{rc}
r_c &= 1 - \frac{\kappa_{wg}}{\tilde{\kappa}(1+\tilde{\eta})} \\
t_u &= -\frac{\sqrt{\kappa_{sc} \kappa_{wg}}}{\tilde{\kappa}} \\
t_c &= t_u/(1+\tilde{\eta}) \\
l_u &= 0 \\
l_c &= \frac{i g \sqrt{\gamma \kappa_{wg}}}{\tilde{\gamma} \tilde{\kappa} (1+\tilde{\eta})}
\end{align}
For an initial atomic state $|\psi_{in}\rangle = \ket{+} = (\ket{u} + \ket{c})/\sqrt{2}$ and a coherent state $\alpha$ in $a_{wg}$, the output state is:
\begin{equation}
\ket{\psi_{out}} = \left(\ket{u,r_u \alpha, t_u \alpha, l_u \alpha} + \ket{c,r_c \alpha, t_c \alpha, l_c \alpha}\right)/\sqrt{2},
\end{equation}
where the output state is  labeled by the atomic state and the three coherent state amplitudes in the output modes $\{a_{wg},\xi_{sc},\xi_{at}\}$. If we detect at least one photon in the mode $a_{wg}$, then the conditional state of the system $\ket{\psi^{cond}_{out}} \sim a_{wg} \ket{\psi_{out}}$ is given by:
\begin{equation}
\ket{\psi^{cond}_{out}} = \frac{1}{\sqrt{\left| r_u \alpha\right|^2 + \left| r_c \alpha \right|^2}} \left(r_u \alpha \ket{u,r_u \alpha, t_u \alpha, l_u \alpha} + r_c \alpha \ket{c,r_c \alpha, t_c \alpha, l_c \alpha} \right).
\end{equation}
Since the residual photons in all three modes are eventually measured or lost to the environment, the state of the system in both the unconditioned and conditioned cases can be described by tracing over all photon modes. This results in the reduced density matrix for the atom:
\begin{align}
\rho_{out} &= \frac{1}{2}\left( \begin{array}{cc} 1 & D \\ D^* & 1\end{array}\right) \\
\rho^{cond}_{out} &= \frac{1}{(|r_u|^2 + |r_c|^2)}\left( \begin{array}{cc} |r_u|^2 & D r_c^*r_u  \\ D^* r_u^*r_c  & |r_c|^2\end{array}\right) \\
D &= \< r_c \alpha, t_c \alpha, l_c \alpha | r_u \alpha, t_u \alpha, l_u \alpha \>
\end{align}
where $D$ is the overlap between the output photonic states, which scales as $e^{-|\alpha|^2}$. Now we calculate two fidelities, quantifying the extent to which we leave the input state untouched without conditioning ($P^{unc}$), and the overlap with the target output state with conditioning ($P^{cond}$). These two quantities are related to the visibility of the green and blue curves in Fig. 4 of the main text, respectively. They are given by:
\begin{align}
P^{uncond} &= \<+|\rho_{out}|+\> = \frac{1}{2}(1+\mathrm{Re}[D]) \\
P^{cond} &= \<-|\rho^{cond}_{out}|-\> = \frac{1}{2(|r_u|^2 + |r_c|^2)} \left(|r_u|^2 + |r_c|^2 - \mathrm{Re}[D r_c^*r_u] \right).\label{eq:Pcond} 
\end{align}
These expressions can be easily generalized to evaluate an overlap with arbitrary atomic state.

To illustrate the combined effect of photon loss and multiphoton excitations, we set $r = r_u = - r_c$. Even in the presence of losses ($k<1$) this case is experimentally achievable by balancing the cavity losses with a finite cooperativity. In this case, the conditional and unconditional fidelities become the same: $P^{uncond} = P^{cond} = \frac{1}{2}(1+D)$. Additionally, if $\eta \gg 1$, $D$ takes a simple form such that:
\begin{equation}
P^{uncond} = P^{cond} = \frac{1}{2}(1+e^{-(1+r^2)|\alpha|^2}). 
\end{equation}
Note that while a lower $\alpha$ results in a higher conditional fidelity, it also decreases the probability of successfully flipping the switch, which scales as
$P = 2 \epsilon \eta_c r^2 / (1+r^2)$, where $\eta_c$ is the total detection efficiency of reflected photons and $\epsilon = P^{cond}-1$ is the error of the gate operation.

For our experimental parameters ($k=0.8$, $\eta\simeq8$ and $|\alpha|^2=0.6$) 
we estimate a gate fidelity from (\ref{eq:Pcond}) of $P^{cond}=0.79$ and $P^{uncond}=0.80$ on resonance, mostly limited by the relatively large $|\alpha|^2$. Additional reduction of the fidelity arises from spontaneous scattering events by the gate photon leaving the atom in a different hyperfine sub-level than $|c\rangle$ ($\sim 10\%$) and from imperfect hyperfine state preparation and readout ($\sim 10\%$) as discussed in section \ref{sect:suppdisc}. 

Finally, implicit in this discussion is the assumption that the reflected $H$- and $V$-polarized photons occupy the same temporal mode, so the photon arrival time does not contain any which-path information for the photon. Since the $V$-polarized interferometer arm is just a mirror, this requires the response time of the atom-cavity system to be instantaneous as well, in the sense that the intracavity field must equilibrate to its steady-state value on a timescale that is short compared to the duration of the input pulse. In our experiment,  the bandwidth of the coupled atom-cavity system is given by $\eta \gamma$.  
Thus, as long as the temporal mode containing the input photon is much longer than the inverse of this rate, the output mode should be identical to the input mode and the gate will operate as expected. In practice, we use gaussian pulses of duration 24 ns (FWHM), which is not quite long enough to fully neglect retardation in the atomic dynamics. We estimate that this introduces imperfections at a level of less than 10\%.

\end{document}